\documentclass [12pt,a4paper]{article}

\usepackage{amsmath,amssymb,amsfonts}
\usepackage[T1]{fontenc}
\usepackage{mathptmx}
\usepackage{flushend}
\usepackage[numbers,sort&compress]{natbib}
\usepackage[pdftex,colorlinks,citecolor=blue,
urlcolor=blue,linkcolor=blue]{hyperref}
\usepackage[pdftex]{graphicx}

\renewcommand{\abstractname}{}\abstractname

\title{Null and Timelike Geodesics Near the Throats of Phantom Scalar Field Wormholes}

\author{\large Ivan  Potashov\thanks{potashov.im@tversu.ru}\,,\, Julia Tchemarina\thanks{chemarina.yv@tversu.ru}\,,\, and Alexander Tsirulev\thanks{tsirulev.an@tversu.ru}}
\date{\small Faculty of Mathematics, Tver State University, Sadovyi per. 35, 170002 Tver, Russia}

\begin{document}

\maketitle

\vspace{-9ex}
\begin{abstract}
We study geodesic motion near the throats of asymptotically flat, static, spherically~symmetric traversable wormholes supported by a self-gravitating minimally coupled phantom scalar field with an arbitrary self-interaction potential. We assume that any such wormhole possesses the reflection symmetry with respect to the throat, and consider only its observable ``right~half''. It turns out that the main features of bound orbits and photon trajectories close to the throats of such wormholes are very different from those near the horizons of black holes. We~distinguish between wormholes of two types, the first and second ones, depending on whether the redshift metric function has a minimum or maximum at the throat. First, it turns out that orbits located near the centre of a wormhole of any type exhibit
retrograde precession, that is, the angle of pericentre precession is negative. Second, in the case of high accretion activity, wormholes of the first type have the innermost stable circular orbit at the throat while those of the second type have the resting-state stable circular orbit in which test particles are at rest at all times. In our study, we have in mind the possibility that the strongly gravitating objects in the centres of galaxies are wormholes, which can be regarded as an alternative to black holes, and the scalar field can be regarded as a realistic model of dark matter surrounding galactic centres. In this connection, we~discuss qualitatively some observational aspects of results obtained in this article.
\end{abstract}

\section{Introduction}

The centres of galaxies are usually recognized to be supermassive black holes~\cite{Goddi2017,Akiyama2018,Abuter2020}, but~in fact we still do not have a completely reliable identification of these objects~\cite{Johannsen2016,BisnKogan2015}. In~this connection, the~natural question arises as to how one may distinguish black holes from other strongly gravitating objects such as naked singularities~\cite{Joshi2014,Shaikh2018}, boson stars~\cite{Macedo2013,Grould2017}, and~wormholes~\cite{BronnikovFabris2006, BronnikovMelnikov2007, Kardashev2007, Dai2019} whose properties may be very closely analogous to those of black holes~\cite{Nandi2006,Tsukamoto2017, Krasnikov2018}. In~the context of astronomical observations, bound~orbits of massive particles and trajectories of photons play a key role in dealing with this question, because~they give very much information concerning the geometry of spacetime near the centres of galaxies~\cite{Mishra2018, Willenborg2018, Stashko2018}. However any interpretation of the observations should be based on a relevant mathematical model of geodesic motion near the central objects. In~particular, dark matter surrounding the centres should be taken into account~\cite{Lee1996,Robles2012, Dokuchaev2015,Cunha2019}.

In this article we study the geodesic motion in the spacetime of a scalar field wormhole, connecting~two asymptotically Schwarzschild universes, using a simple but completely self-consistent approach, in~which dark matter is modelled by a scalar field~\cite{Matos2004, UrenaLopez2019} while effects of rotation are not taken into account; then the problem can be analytically tractable. This approach allows us to explore some unevident features of the trajectories of massive and massless test particles in such spacetimes; these features are usually hidden in qualitative models based on computer simulations. Other useful fully analytical models are based on the Kerr-like configurations, which, however, do not take account of dark matter~\cite{Johannsen2016, Gourgoulhon2018, Dokuchaev2019}. We will consider the simplest case of a spherically symmetric, static, traversable wormhole supported by a nonlinear self-gravitating minimally coupled phantom (in another terminology, ghost) scalar field with an arbitrary self-interaction potential. We also will assume that the wormhole connects two asymptotically flat regions of spacetime, has a single throat, and~possesses the reflectional symmetry with respect to the throat. Similar models of wormholes are studied in references~\cite{Bronnikov2017, Kratovitch2017, Ovgun2019}, and~more general models are considered in references~\cite{BronnikovChernakova2007, BronnikovSushkov2010, Bronnikov2018, Bronnikov2019}. Distinctive~features of geodesic motion near the throat of a phantom scalar field wormhole can be used to identify observationally the central objects in galaxies, and~also to test general relativity in a strong gravitational regime in order to distinguish between some dark matter models and the hypothetical fifth force~\cite{Hees2017}.

The paper is organized as follows. In~Section~\ref{ScalarfieldWHs}, writing the metric of static spherially symmetric spacetime in the so-called quasiglobal coordinates, we derive and consider the main features and properties of the wormhole solutions to the Einstein-Klein-Gordon equations. Section~\ref{TimelikeNullG} contains the mathematical description of geodesic motion in the  spacetime of a static traversable scalar field wormhole which possesses the reflectional symmetry with respect to its throat. In~this section, we~also classify such wormholes into two qualitatively different types according to the form of the metric redshift function. We derive and analyse the behaviour of null and timelike geodesics for two analytical wormhole solutions in Section~\ref{Examples}. Section~\ref{Disscussion} provides a brief discussion and final remarks concerning some observational aspects of geodesic motion near wormhole throats that distinguish them from other strongly gravitating objects.

The signs of the Riemann and Ricci tensor fields are defined as $R^i_{jkl}=\partial_k\Gamma^i_{jl}-\ldots$ and $R_{jl}=R^i_{jil}$, respectively. Throughout the article we adopt the metric signature $(+,-,-,-)$ and use the geometric system of units in which $G=c=1$.

\section{Scalar Field~Wormholes} \label{ScalarfieldWHs}

This section contains some mathematical preliminaries and relevant statements (for the most part known earlier) for studying the geodesic motion in the background of a wormhole supported by a phantom scalar field. We consider some basic properties of static spherically symmetric and asymptotically flat wormholes using the so-called inverse problem method for a self-gravitating scalar field. We begin with the action
\begin{equation}\label{action}
    S=\frac{1}{8\pi}\int\left(-\frac{1}{2}R+ \varepsilon\langle d\phi,d\phi\rangle-2V(\phi)\right) \sqrt[]{|g|}\,d^{\,4}x\,,
\end{equation}
where $R$ is the scalar curvature, $V(\phi)$ is a self-interaction potential of a nonlinear scalar field $\phi$, the~angle brackets denote the pointwise scalar product induced by a spacetime metric $g$, and~$\varepsilon=\pm1$. Note that $\varepsilon=-1$ corresponds to the case of a phantom scalar field. The~negative sign of the kinetic term in the action~(\ref{action}) is a necessary condition for the existence of wormhole~solutions.

For a static spherically symmetric wormhole spacetime, there exist the most natural coordinates, namely the quasiglobal coordinates, in~which the metric has the form
\begin{equation}\label{metric}
ds^2=Adt^2-\frac{\,dr^2}{\,A}-C^{2} \left(d\theta^2+\sin^{2}\theta\,d\varphi^2\right),
\end{equation}
where the metric functions $A$ and $C$, as~well as the field $\phi$, are functions only of the radial coordinate $r$ with the range from $-\infty$ to $\infty$. These functions are assumed to be of class $\mathcal{C}^2$ everywhere and to be analytical at $r=0$. Without~loss of generality, from~now on we set $r=0$ on the throat. We restrict our attention to traversable wormholes that are symmetric with respect to the throat. This means that the metric functions are positive everywhere and invariant under the reflection:
\begin{equation}\label{cond1}
A(r)>0,\quad A(-r)=A(r),\qquad  C(r)>0,\quad C(-r)=C(r), \qquad r\in(-\infty,\infty).
\end{equation}

The asymptotic conditions are
\begin{equation}\label{cond2}
A=1-\frac{2M}{|r|}+O\!\left(|r|^{-2}\right), \quad C=|r|+\alpha+O\!\left(|r|^{-1}\right),\quad r\rightarrow\pm\infty,
\end{equation}
where $M$ is the Schwarzschild gravitational mass as measured by a distant~observer.

In the orthonormal basis associated with the metric~(\ref{metric}), the~independent Einstein-Klein-Gordon field equations for the action~(\ref{action}) are
\begin{eqnarray}\label{00}
-2A\frac{\,C''}{C} - A'\,\frac{\,C'}{C} - A\frac{{\,C'}^2}{C^2} + \frac{1}{\,C^2} &=& \varepsilon{}A{\phi'}^2 + 2V\,, \\
\label{11}
A'\frac{\,C'}{C} + A\frac{{\,C'}^2}{C^2} - \frac{1}{\,C^2} &=& \varepsilon{}A{\phi'}^2 -2V\,, \\
\label{KG}
-A\phi''- \phi'\!\left(A'+ 2A\frac{\,C'}{C}\right)+ \varepsilon\frac{dV}{d\phi} &=& 0\,,
\end{eqnarray}
where a prime denotes differentiation with respect to the radial coordinate $r$. Both for classical and phantom scalar fields, there exists a special algorithm, the~so-called 'inverse problem method for self-gravitating scalar field', for~constructing one-parameter families of static spherically symmetric solutions to the Einstein-Klein-Gordon equations. This method developed in references~\cite{Lechtenfeld1995, BronnikovShikin2002, Nikonov2008, Tchemarina2009, Azreg2010} allows us to examine the problem, in~some sense, for~all admissible self-interaction potentials at the same time. We will use a variant of the inverse problem method in which a 'general' solution of Equations~(\ref{00})--(\ref{KG}) is represented by simple quadratures~\cite{Solovyev2012}.

In the case under consideration (namely, $\phi$ is a phantom field with $\varepsilon=-1$, and~the \mbox{conditions~(\ref{cond1}) and~(\ref{cond2})} hold), the~quadratures have the form
\begin{equation}\label{A}
A\,=\,2C^{2}\!\int_{r}^{\infty}\! \frac{\,r\,}{C^4}\,dr,
\end{equation}
\begin{equation}\label{phi'}
{\phi'}^2=\, C''/C,
\end{equation}
\begin{equation}\label{Vr}
\widetilde{V}(r)= \big(1-3A{C'}^2- ACC''+ 2rC'\!/C\big)\big/\big(2C^2\big),
\end{equation}
\begin{equation}\label{Vphi}
V(\phi)= \widetilde{V}\big(r(\phi)\big).
\end{equation}

It is not hard to verify that expressions~(\ref{A})--(\ref{Vr}) reduce Equations~(\ref{00})--(\ref{KG}) to identities. Moreover,~one can formulate the converse as follows: if $(A,C,\phi)$ is a solution of Equations~(\ref{00})--(\ref{KG}) for a given potential $V$, and~if the conditions~(\ref{cond1}) and~(\ref{cond2}) hold, then the relations~(\ref{A})--(\ref{Vphi}) are true. From~Equation~(\ref{phi'}), which is the sum of~(\ref{00}) and (\ref{11}) divided by $-2A$, one sees that $C''(r)\geq0$ everywhere. It means that if $C''(0)\neq0$, then $\phi(r)$ is necessarily an odd function. If~$C''(0)=0$, then the field can be thought of either as an even or odd function. In~order to avoid multiple repetitions\label{ppp} and for definiteness, we will consider analytically only the right half of a wormhole (that is, the~region $r\geq0$) and extend $\phi(r)$ to the left half as an odd~function.

A convenient mathematical technique for constructing wormhole solutions consists of the following steps: first, one should specify an appropriate function $C(r)$; second, it is necessary to calculate the metric function $A(r)$, the~field $\phi(r)$, and~the potential $\widetilde{V}(r)$ as a function of the radial coordinate; finally, taking into account that $\widetilde{V}(r)$ is necessarily an even function, the~self-interaction potential $V(\phi)$ can be restored by the formula~(\ref{Vphi}). Another way of constructing the solutions is to specify a monotonic field function $\phi(r)$, and~then to solve the equation $C''-{\phi'}^2C=0$ for $C(r)$. Note also that the condition~(\ref{cond2}) and the quadrature~(\ref{A}) allow us to write the metric~(\ref{metric}) at infinity in the Schwarzschild coordinates; we find that the Schwarzschild mass is uniquely determined by the constant $\alpha$, and~is given by
\begin{equation}\label{M}
M=\alpha/3.
\end{equation}

We will suppose below that $M>0$ because the gravitational field of a zero mass wormhole does not approach to the Newtonian limit at infinity. For~example, if~$C=r+p/r+q/r^2+o\big(r^{-2}\big)\;(q\neq0)$ as~$r\rightarrow+\infty$, then one find from~(\ref{A}) that $A=1+2q/(5r^3)+O\big(r^{-4}\big)$. It shows that such a wormhole possesses a non-Newtonian gravitational potential \big(namely, $\Phi=q/(5r^3)$\big). Moreover, if~$q>0$, the~gravitational force is~repulsive.

As a consequence of the condition~(\ref{cond1}), the~expansions of the metric functions in the vicinity of the throat have the form
\begin{equation}\label{AC-expan}
A\,=\,A_0+ \frac{1}{2}A''(0)r^2+ \frac{1}{24}A^{(4)}(0)r^4+\ldots,\qquad C\,=\,C_0+ \frac{1}{2}C''(0)r^2+ \frac{1}{24}C^{(4)}(0)r^4+\ldots\,.
\end{equation}

Directly from~(\ref{A}) and~(\ref{AC-expan}) we find that
\begin{equation}\label{A-deriv}
A_0= 2C_0^{2}\!\int_{0}^{\infty}\! \frac{\,r\,}{C^4}\,dr,\quad
A''(0)= \frac{2}{C_0^2}\big(A_0C_0C''(0)-1\big),\quad
A^{(4)}(0)= \frac{2A_0}{C_0^2}\Big(C_0C^{(4)}+3[C''(0)]^2\Big).
\end{equation}

It follows from~(\ref{cond2}),~(\ref{A}), and~(\ref{M}) that $A'=2Mr^{-2}+O(r^{-3})$ as $r\rightarrow+\infty$, that is, $A'>0$ in some neighbourhood of the (positive) spatial infinity, and~therefore $A(r)$~reaches its minimum value at some point $r=r_{min}\geq0$ in a coordinate neighbourhood of the throat; in the case $r_{min}>0$, we will assume that $A(r)$~decreases in the region~$(0,r_{min})$. These arguments imply that there are two geometrically and observationally different types of wormhole configurations. In~wormhole spacetimes of the first type (respectively, second type), the~metric function $A(r)$ reaches a minimum (respectively, maximum) at the throat. The~minimum (respectively, maximum) at $r=0$ takes place if and only if there exists an integer $n\geq1$ such that
\begin{equation}\label{A2nth}
A^{(2k)}(0)=0 \;\; (k=1,\,\ldots,\,n-1), \qquad A^{(2n)}(0)>0\;\;(\text{respectively, }A^{(2n)}(0)<0).
\end{equation}

Note that we exclude from our consideration the special case $A\equiv1$ represented by the unique (apart from flat spacetime) Ellis--Bronnikov solution~\cite{Ellis1973, Bronnikov1973, Ellis1974, Ellis1979} with $\phi(r)=\arctan\!\big(r/C_0\big)$, $V(\phi)=0$, and~$C(r)=\big(r^2+C_0^2\big)^{1/2}$;\, this solution represents a traversable wormhole which is geodesically complete but admits no stable bound orbits at~all.

In what follows, we take, for~the sake of brevity,  $n=1$ in~(\ref{A2nth}). This restriction is motivated by numerical experiments which have shown that the qualitative behaviour of geodesics for~$n=1$ is essentially the same as for~$n\geq2$. Then it follows directly from~(\ref{A-deriv}) that the conditions~(\ref{A2nth}) can be rewritten in the form
\begin{equation}\label{A''0}
A_0C_0C''(0)-1>0\;\;(\text{for the first type}),\quad
A_0C_0C''(0)-1<0\;\;(\text{for the second type}).
\end{equation}

In particular, a~wormhole is of the second type if~$C''(0)=0$. Note that the constant $C_0$ is the size of the throat, and~$A_0$ can be interpreted either as the potential barrier height (if $A''(0)<0$) or as the potential well depth (if $A''(0)>0$) for test particles falling radially toward the~throat.

The geometrical system of units does not fix a unit of length. On~the other hand, the~geodesic structure of spacetime is scale invariant at the classical level, and~Equations~(\ref{00})--(\ref{KG}) and the quadratures~(\ref{A})--(\ref{Vphi}) are also invariant under the scale transformations
\begin{equation}
r\rightarrow r/\lambda,\quad M\rightarrow M/\lambda,\quad V\rightarrow \lambda^2V,\quad\;\;\lambda>0.\nonumber
\end{equation}

Therefore, we can use an arbitrary unit of length. By~applying $\lambda=M$ in this transformation, we~can take, as~it is usually done in general relativity, the~Schwarzschild mass of a configuration as the current unit of length. Thus, without~loss of generality, we suppose everywhere below that $M=1$.

\section{Timelike and Null~Geodesics}
\label{TimelikeNullG}

In all stationary spherically symmetric spacetimes we have the conserved energy and angular momentum of a test particle. Together with the constancy of the norm of the four-velocity, this implies the existence of three integrals of motion~\cite{Chandrasekhar1998}. Supposing that a particle moves in the equatorial plane, these integrals for the metric~(\ref{metric}) can be written in the form
\begin{equation}\label{JE1}
C^2\frac{d\varphi}{ds}= J\,,\quad A\frac{dt}{ds}= E\,, \quad \left(\!\frac{dr}{ds}\!\right)^{\!\!2}=  E^2- V_{eff},
\end{equation}
\begin{equation}\label{Veff}
 V_{eff}= A\!\left(\mathrm{k}+\dfrac{\,J^2}{\,C^2}\right),
\end{equation}
where $J$ is the specific angular momentum, and~$E$ is the specific energy of a test particle. For~massive test particles, $\mathrm{k}=1$ and $s$ is the proper time, while for massless ones $\mathrm{k}=0$ and $s$ is an arbitrary affine parameter. The~effective potential $V_{eff}$ is obviously an even function, and~$V_{eff}\rightarrow1$ ($V_{eff}\rightarrow0$) for massive particles (respectively, for~massless ones) as $r\rightarrow\infty$.

From the point of view of a distant observer at rest with respect  to the centre, the~angular velocity of a test particle along its trajectory is given by $$\omega=\dfrac{d\varphi}{dt}= \frac{J}{E}\:\!\dfrac{A}{C^2}$$
both for null and timelike geodesics. Note that $\omega=0$ for massive particles in the resting-state stable circular orbit where $J=0$.

\subsection{Timelike~Geodesics}
\label{TimelikeNullG-1}

Let $C$ be specified and $A$ be found from~(\ref{A}). Then any timelike geodesic is determined by its arbitrary point and the parameters $J$ and $E$. We are primarily interested in circular orbits of massive particles near the throats of wormholes, so that we set $\mathrm{k}=1$ in this section. From~(\ref{Veff}) we have
\begin{equation}\label{Veff'}
V_{eff}'(r,J)= A'+J^2{\left(\!\frac{A}{C^2}\!\right)\!}',
\end{equation}
where a prime denotes differentiation with respect to $r$. Circular orbits correspond to solutions of the equation $V_{eff}'(r,J)=0$. For~circular orbits in the region $r>0$, the~$r$-component of the geodesic equation gives
\begin{equation}\label{r-comp} A'\!\left(\frac{dt}{ds}\right)^{\!\!2}- \big(C^2\big)'\! \left(\frac{d\varphi}{ds}\right)^{\!\!2}=0.
\end{equation}

This relation together with~(\ref{JE1}) allows us to express the first two integrals of (circular) motion as functions only of the radial coordinate:
\begin{equation}\label{}
J^2=C^2\frac{A'}{A} \left(2\frac{C'}{C}-\frac{A'}{A}\right)^{\!\!-1}, \quad
E^2=2A\frac{C'}{C} \left(2\frac{C'}{C}-\frac{A'}{A}\right)^{\!\!-1}.
\nonumber
\end{equation}

On the right side of a wormhole, timelike circular orbits may exist only in the region where the conditions $A'>0$ and $2C'/C-A'/A>0$ hold; the latter is equivalent to
\begin{equation}\label{C^2/A}
\big(A/C^2\big)'<0.
\end{equation}

It follows directly from~(\ref{A}) that $\big(A/C^2\big)'<0$ for all $r>0$. In~wormholes of the first type, the~condition $A'>0$ holds for all $r>0$. However, in~wormholes of the second type, this condition holds only in the region $r>r_{min}$. On~the throat, the~limit equality $\big(A/C^2\big)'=0$ is identically satisfied due to the conditions~(\ref{AC-expan}).

Massive particles, moving along circular orbits close to the throat, may or may not form an accretion disk, which in turn may or may not have the innermost stable circular orbit. Let us first consider wormholes of the first type for which $A''(0)>0$, $A'(0)=C'(0)=0$; the latter implies $V_{eff}'=0$. Taking into account the second equality in~(\ref{A-deriv}), we find from~(\ref{Veff}) that
\begin{equation}\label{rJ}
V_{eff}''(0,J)= A''(0)-\frac{2J^2}{C_0^4},\qquad V_{eff}''(0,J)=0\;\Rightarrow\; J=J_0\equiv C_0^2\frac{\sqrt{A''(0)}}{2}.
\end{equation}

The value $J_0=C_0^2\sqrt{A''(0)/2}\;$ separates stable and unstable orbits located at the throat. Such an orbit is stable if $J<J_0$, and, moreover, particles in orbits with $J=0$ are at rest. If~$J>J_0$, then orbits located at $r=0$ are unstable, but~for each $r>0$ there exists a stable orbit. Indeed, if~$V_{eff}''(0)<0$ and $V_{eff}'>0$ at infinity, the~effective potential has a minimum at some point $r=r_*>0$ at which $V_{eff}''>0$. On~the other hand, it follows from~(\ref{Veff'}) and~(\ref{C^2/A}) that there exists a unique solution, say $r_*(J)$, of~the equation $V_{eff}'(r,J)= 0$ for each $J>J_0$. Moreover, we have the inequality
\begin{equation}\label{r*J}
V_{eff}'(r_*,J)= 0\quad\Rightarrow \quad\frac{dr_*}{dJ}=- \frac{1}{V_{eff}''} \frac{dV_{eff}'}{dJ} \Big|_{r=r_*}= -\frac{2J}{V_{\!eff}''} {\left(\!\frac{A}{C^2}\!\right)\!}' \Big|_{r=r_*}>0,
\end{equation}
which shows that the coordinate radius $r_*$ of the stable circular orbit increases monotonically as $J$ increases. For~circular orbits close to the throat this fact is not self-evident, because~in this region the spacetime geometry of wormholes is radically different from the geometry of black~holes.

These arguments remain in force for wormholes of the second type, but~in this case stable circular orbits exist only in the region $r\geq{}r_{min}>0$, and~particles are at rest ($J=0$) in the orbit with $r=r_{min}$, where $r_{min}$ is defined after the expressions~(\ref{A-deriv}). Thus, an~important result of our analysis is that a scalar field wormhole can have a continuous accretion disk which stretches from some $r=r_{min}\geq0$ to the orbit of a source of diffuse matter revolving around the wormhole. For~definiteness, we will refer to any orbit at $r=r_{min}>0$ as the resting-state stable circular orbit: its difference from the innermost stable circular orbit ($r_{min}=0$) is that particles in the resting-state stable circular orbits are at rest with respect to the throat or have an observationally negligible velocity. The~radius of the resting-state stable circular orbit equals $C_0$ for a wormhole of the first type, while for that of the second type it can be much greater than $C_0$.

Other important astrometric characteristics of bound orbits in the central regions are deviations from the vacuum (Schwarzschild) orbital form. Nowadays the most simple and informative observational data are provided by the precession of pericentres of stellar  orbits located close to the centres. We can express the angle of precession $\Delta\varphi$ of a bound orbit directly from~(\ref{JE1}) as
\begin{equation}\label{precess}
\Delta\varphi\,=\,\varphi_{osc}-2\pi, \qquad
\varphi_{osc}=2J\!\int \limits_{\,r_{p}}^{\:r_{a}}\! \frac{dr} {\,C^2\sqrt{E^2-V_{eff}\,}},
\end{equation}
where $r_{p}$ and $r_{a}$ are the pericentre and apocentre radii, respectively. In~other words, they are solutions of the equation $E^2-V_{eff}=0$ such that $r_{p}< r_{\scriptscriptstyle J}< r_{a}$, where $r_{\scriptscriptstyle J}$ is a (global or local) minimum of $V_{eff}(r,J)$, and~there are no other solutions in the interval $(r_{p},r_{a})$. Thus, a~bound orbit of the general type oscillates near a stable circular orbit (an oscillation is the motion from pericentre to apocentre and back) and $\varphi_{osc}$ is the polar angle which has been passed around the centre between any two successive pericentre points of the orbit. The~relativistic precession of pericentres of bound orbits is considered in~\cite{Grould2017, Dokuchaev2015, Hees2017, Zakharov2012, Meyer2012, Potashov2019, Potashov2019a} both from a purely theoretical point of view and in the context of observations of S-stars in the Galactic~Centre.

\subsection{Null~Geodesics}

For null geodesics $\mathrm{k}=0$, so that~(\ref{JE1}) and~(\ref{Veff}) can be rewritten in the form
\begin{equation}\label{JE0}
C^2\frac{d\varphi}{d\lambda}= b\,,\quad A\frac{dt}{d\lambda}= 1\,, \quad \left(\!\frac{dr}{d\lambda}\!\right)^{\!\!2}=  1-b^2\frac{A}{C^2},
\end{equation}
where $\lambda=sE$ is a new affine parameter, and~$b=J/E$ is the impact parameter. Thus, the~behaviour of a null geodesic is determined only by its arbitrary point and the unique, as~opposed to a timelike geodesic, parameter $b$. We will not consider the trivial case $b=0$ corresponding to a radially falling photon. The~spatial trajectory of a photon satisfies the differential equation
\begin{equation}\label{photon-trajectory}
\frac{dr}{d\varphi}= \pm C^2 \sqrt{\frac{1}{b^2}-\frac{A}{C^2}}.
\end{equation}

An obvious necessary condition for the existence of a null geodesic with a given $b$ is
\begin{equation}\label{}
\frac{1}{b^2}\geq\frac{A}{C^2}, \nonumber
\end{equation}
where the equality holds only at the turning point (the point of closest approach to the throat) or in an unstable circular orbit at the throat. Another important condition for null geodesics also follows directly from~(\ref{JE0}): the trajectory of a photon crosses the throat if and only if
\begin{equation}\label{leave}
\frac{1}{b^2}\geq\frac{A_0}{C_0^2}= 2\!\int_{0}^{\infty}\! \frac{\,r\,}{C^4}\,dr.
\end{equation}

If this condition holds, then the photon leaves forever, in~contrast to bound timelike geodesics, the~region $r<0$.

\section{Two Examples of~Wormholes}
\label{Examples}

\begin{figure}[h]
  \centering
  \begin{minipage}{0.45\textwidth}
    \includegraphics[width=\textwidth]{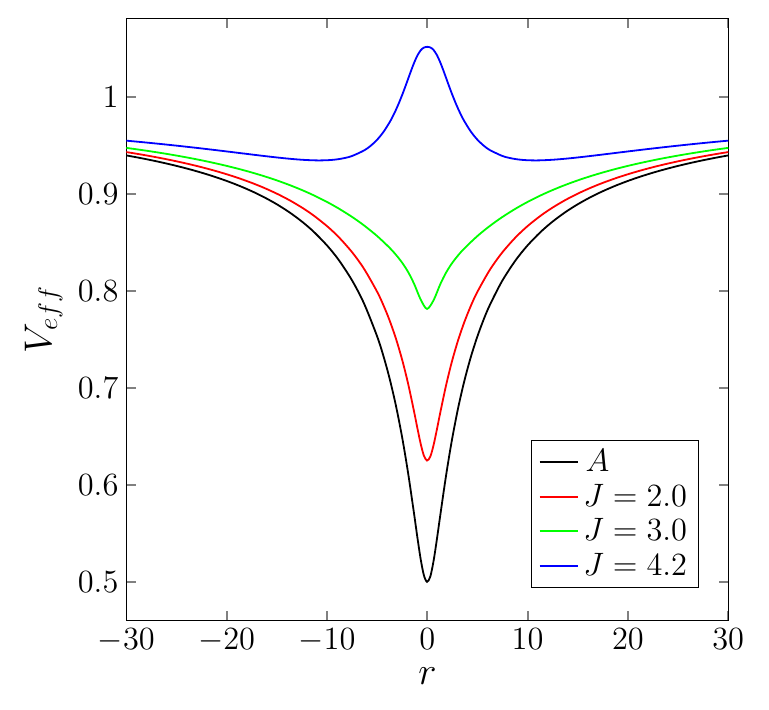}
  \end{minipage}\quad
\begin{minipage}{0.45\textwidth}
    \includegraphics[width=\textwidth]{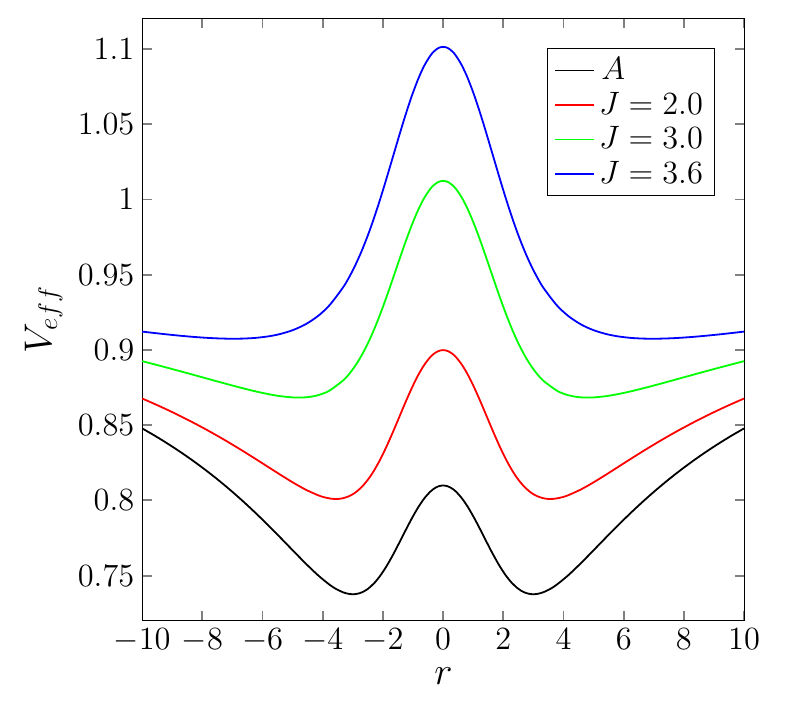}
  \end{minipage}
  \caption{The effective potentials $V_{eff}(r,J)$ of massive particles and (if $J=0$) the metric function $A(r)=V_{eff}(r,0)$ for the ansatz~(\ref{Ca}): $a=1$ (\textbf{left panel}) and $a=9$ (\textbf{right panel}).}
  \label{F1}
\end{figure}

In this section we explore two examples of static wormhole solutions that have not been considered in the literature earlier. They are based on the ansatz
\begin{equation}\label{Ca}
C=\left(r^4+2r^2+a^2\right)^{1/4}+3,\quad a\geq1,
\end{equation}
where $a$ is the parameter of 'intensity' of the scalar field. In~accordance with the conditions~(\ref{cond2}),~(\ref{M}), and~the remark at the end of Section~\ref{ScalarfieldWHs}, all wormholes in the family~(\ref{Ca}) are of unit mass, $M=1$. For~simplicity we will consider wormholes with $a=1$ and $a=9$. For~these values of $a$, the~metric function $A(r)$ can be obtained in terms of elementary functions and elliptic functions, respectively, but~we have not written the explicit expressions for $A(r)$ because of their cumbersome forms. The~self-interaction potential $\widetilde{V}(r)$ is partially negative (in the region $(p,\infty)$, $p>0$) for all $a\geq1$, and~the field $\phi(r)$ is an odd function. The~effective potentials $V_{eff}(r,J)$ of timelike geodesics and the function $A(r)$ are shown in Figure~\ref{F1}. Below~in this section we will need an explicit expression for the unique solution $J_*$ of the equation $V_{eff}(0,J)=1$; from~(\ref{Veff}) we find
\begin{equation}\label{J*}
J_*=C_0\sqrt{1/A_0-1\,}\,.
\end{equation}

\begin{figure}[h]
  \centering
  \begin{minipage}{0.45\textwidth}
    \includegraphics[width=\textwidth]{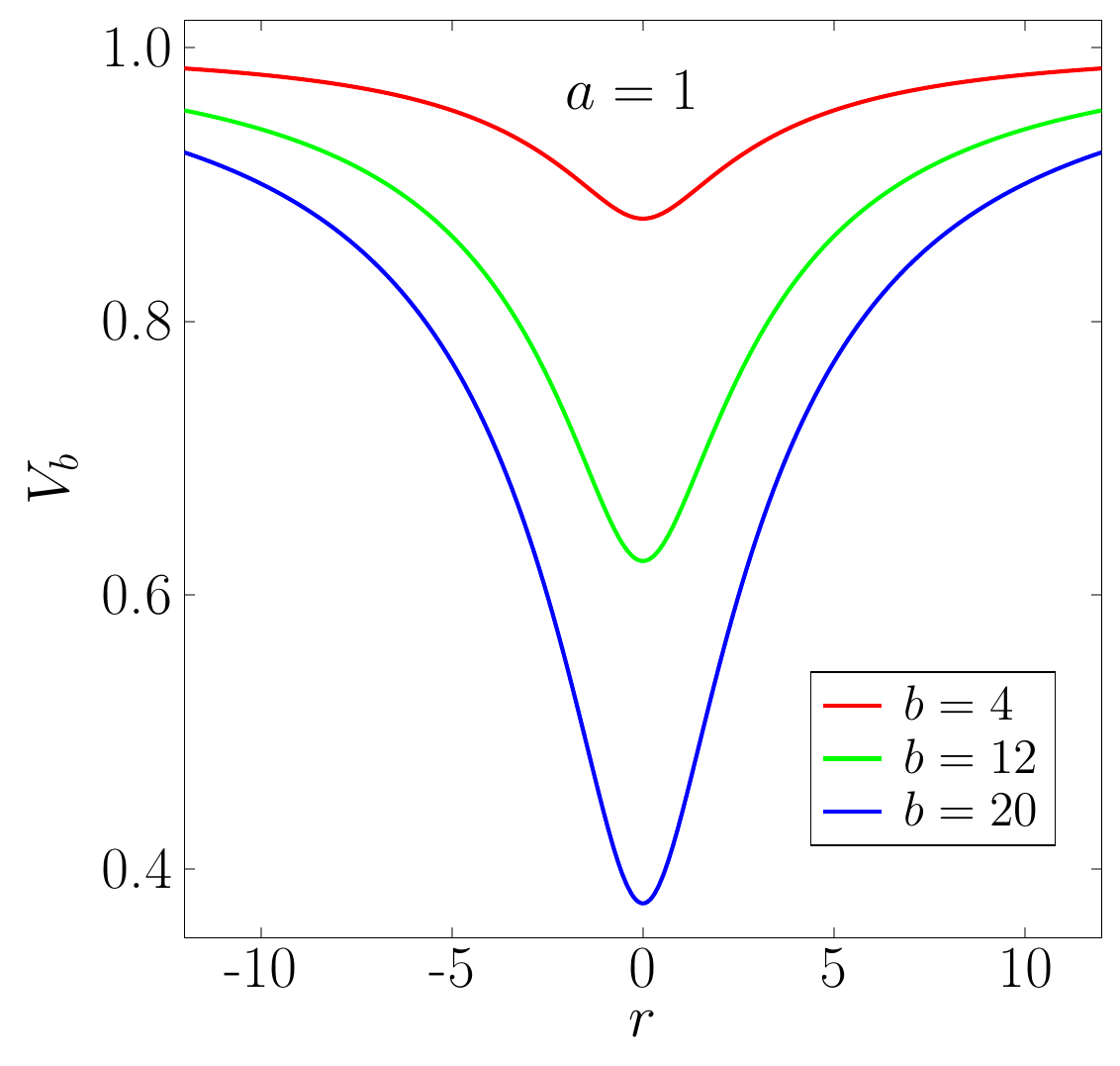}
  \end{minipage}\quad
\begin{minipage}{0.45\textwidth}
    \includegraphics[width=\textwidth]{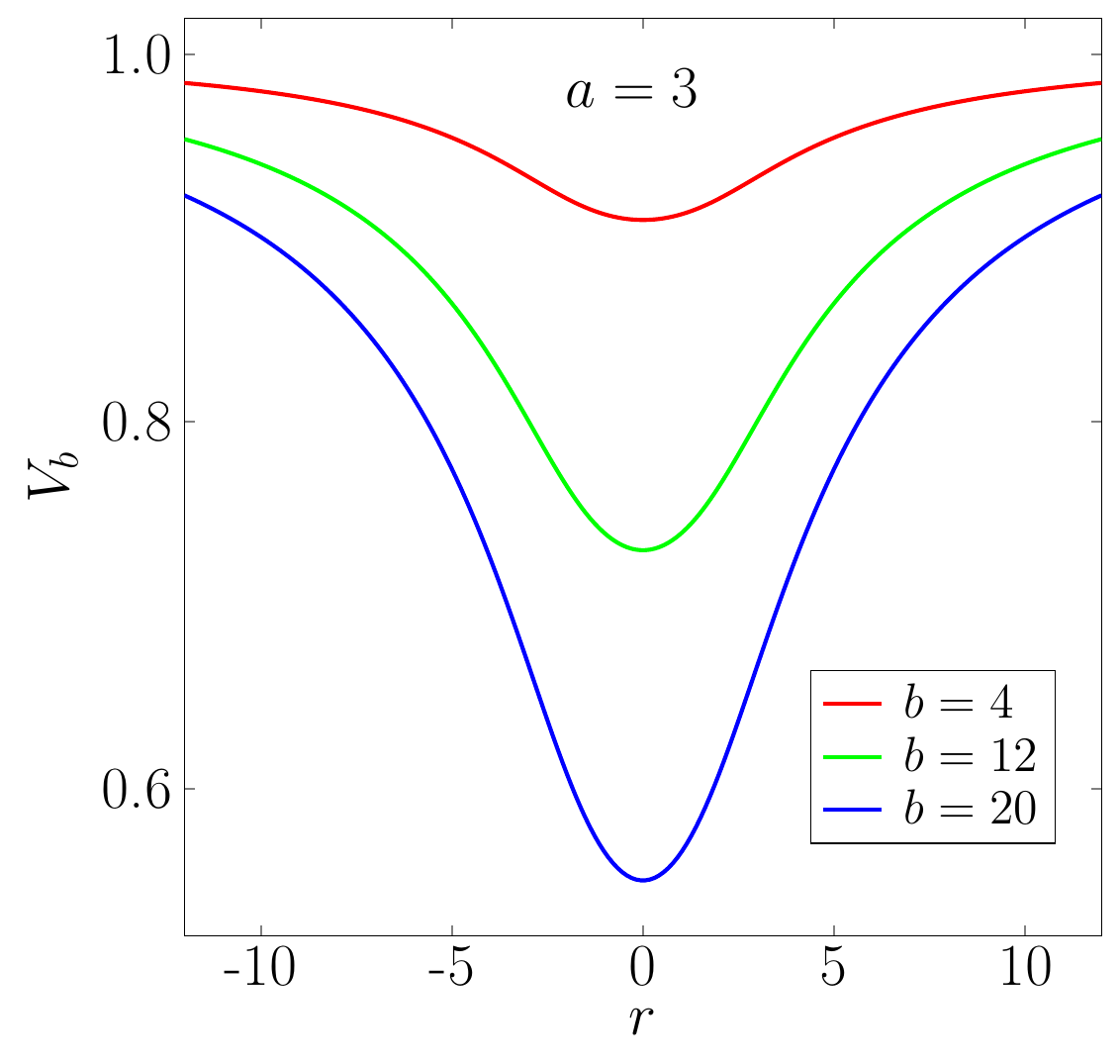}
  \end{minipage}
  \caption{The effective potentials $V_{b}(r,b)=1-b\big(A/C^2\big)$ of massless particles for the ansatz~(\ref{Ca}): $a=1$ (\textbf{left panel}) and $a=9$ (\textbf{right panel}).}
  \label{Fph}
\end{figure}

It should be stressed that bound orbits can exist only in the region where $A<1$. Note also that we consider examples whose types can be clearly recognized: for $a=1$ and $a=9$, the~wormholes are of the first and second types, respectively. The~effective potentials $V_{b}(r,b)=1-b\big(A/C^2\big)$ of null geodesics are shown in Figure~\ref{Fph}.

\unskip

\unskip

\subsection{Trajectories of Timelike~Geodesics}

In our analysis of bound orbits around the throat, we are interested in two the most important observational parameters: the angle of pericentre precession and the radius of the  innermost (or~resting-state) stable circular orbit of the accretion disk. Both the characteristics crucially depend on the type of~wormhole.

\begin{figure}[h]
  \centering
  \begin{minipage}{0.45\textwidth}
    \includegraphics[width=\textwidth]{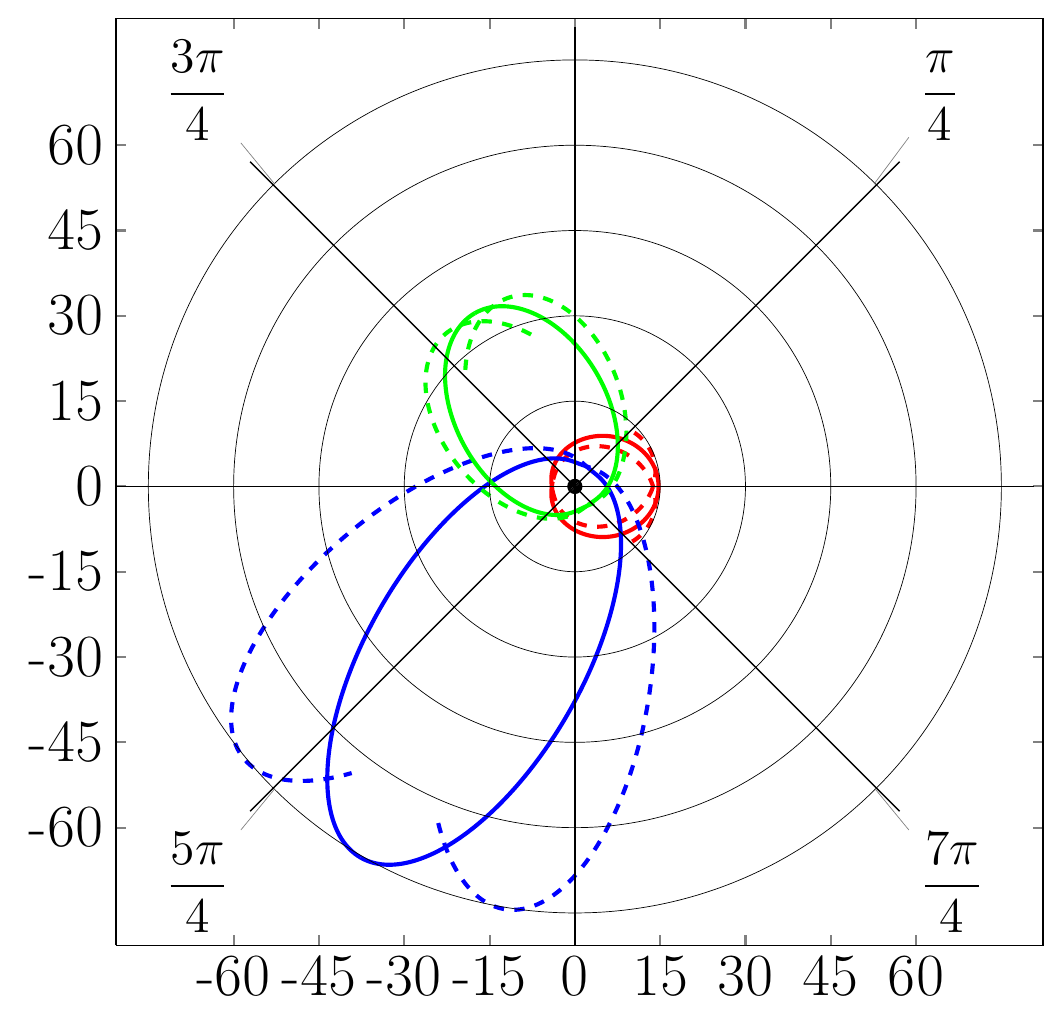}
  \end{minipage}\quad
\begin{minipage}{0.45\textwidth}
    \includegraphics[width=\textwidth]{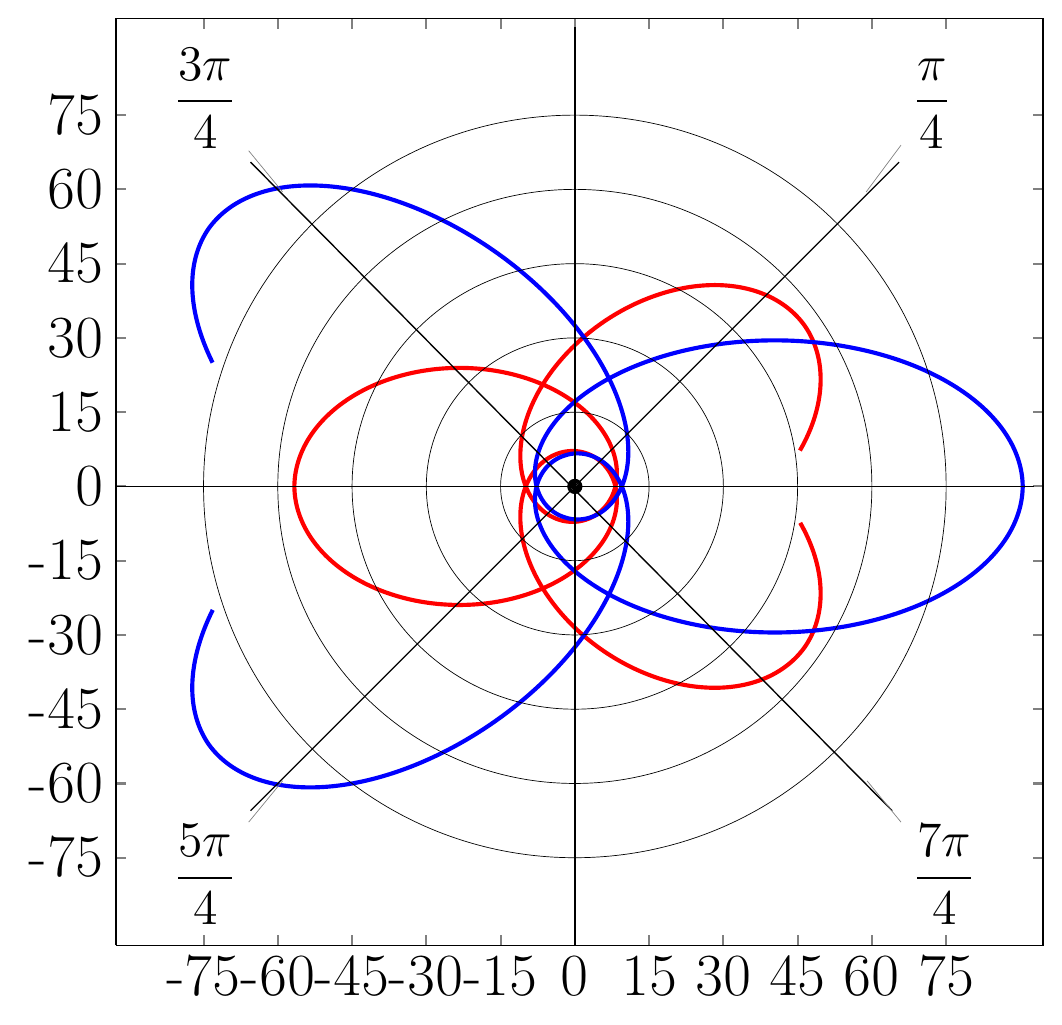}
  \end{minipage}
  \caption{The shapes of some precessing trajectories of massive test particles for ansatz~(\ref{Ca}) with $a=1$. \textbf{Left panel} ($J=3$): $E^2=0.90$, $\Delta\varphi=0.31$~(red line), $E^2=0.95$, $\Delta\varphi=-0.18$~(green line), $E^2=0.975$, $\Delta\varphi=-0.36$~(blue line). The~dashed parts of lines represent the 'hidden' parts of trajectories located in the region $r<0$. \textbf{Right panel} ($J=4.2$): $E^2=0.97$, $\Delta\varphi=2.52$~(red line), $E^2=0.98$, $\Delta\varphi=2.70$~(blue~line).}
  \label{F2}
\end{figure}

\begin{figure}[h]
  \centering
  \begin{minipage}{0.45\textwidth}
    \includegraphics[width=\textwidth]{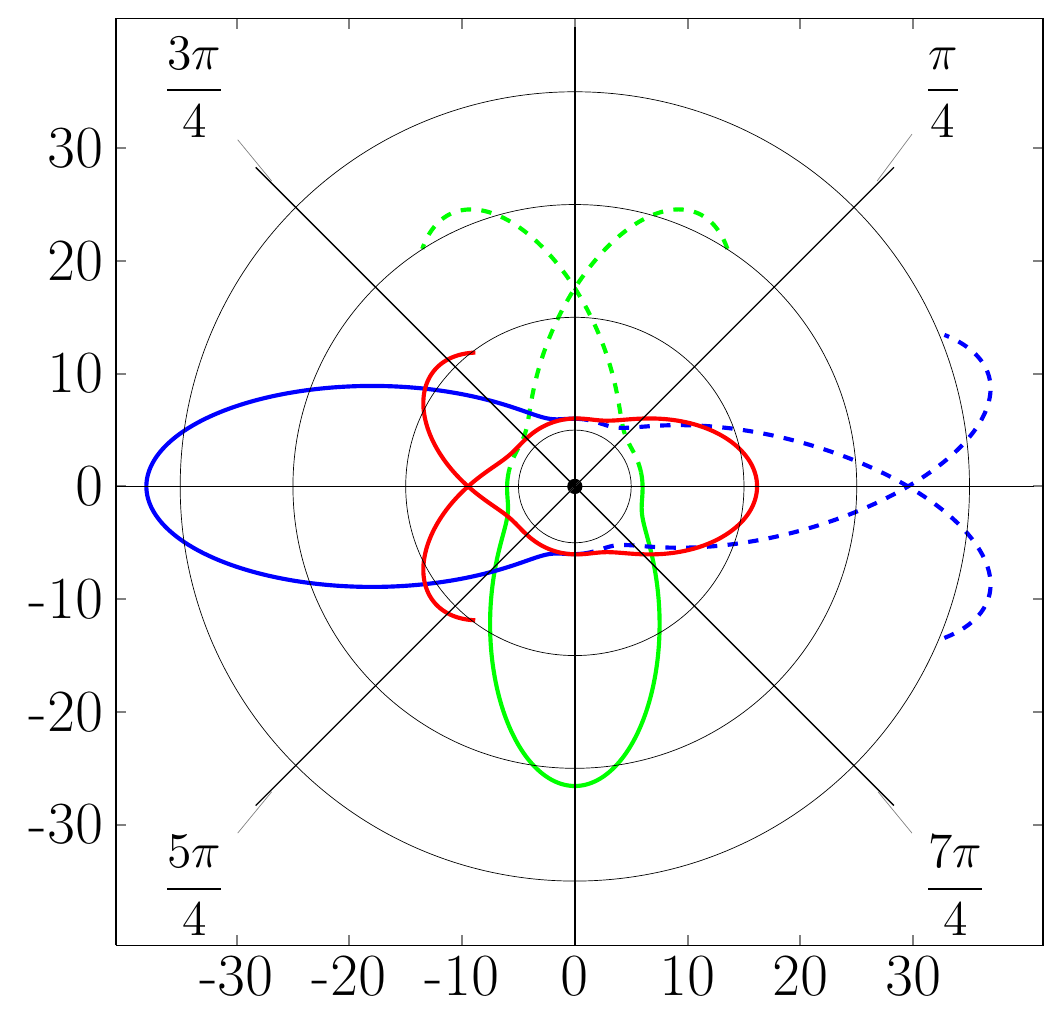}
  \end{minipage}\quad
\begin{minipage}{0.45\textwidth}
    \includegraphics[width=\textwidth]{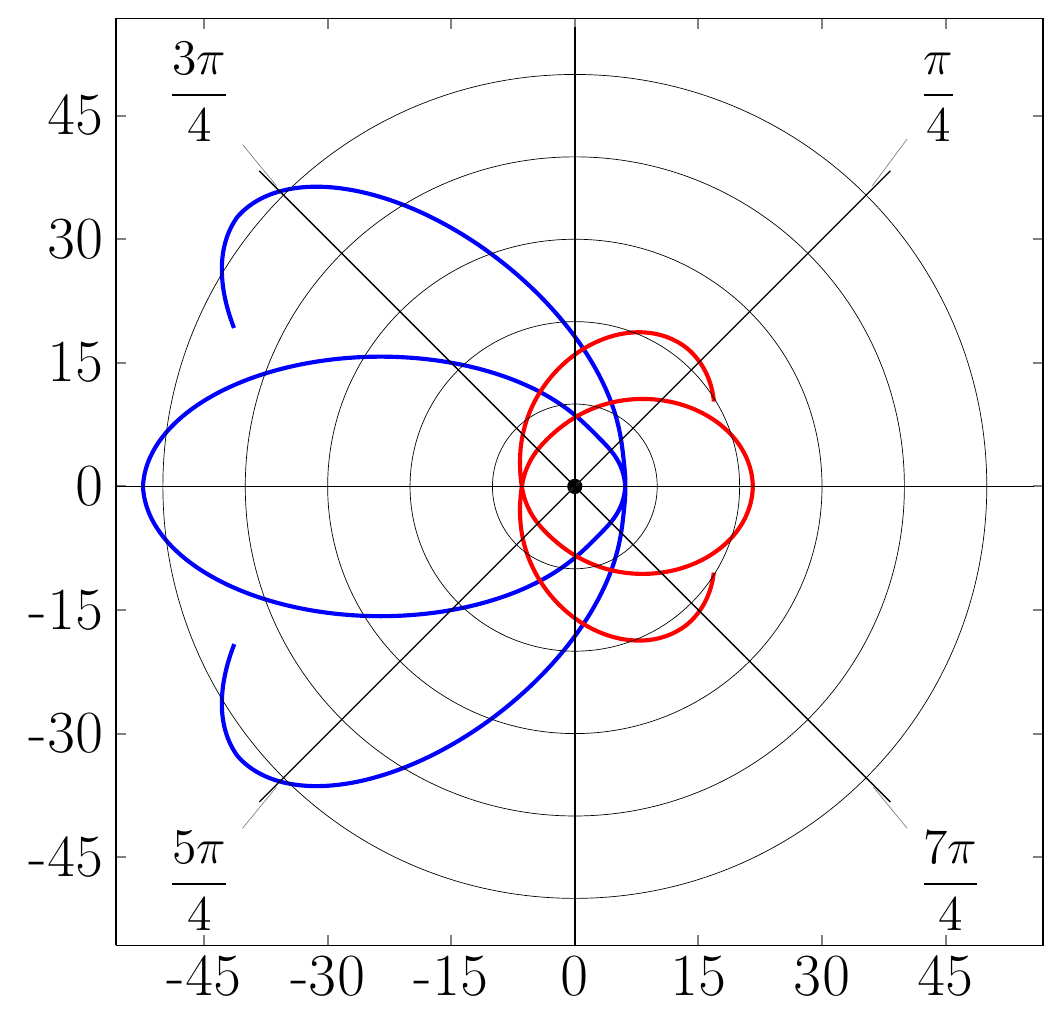}
  \end{minipage}
    \caption{The shapes of some precessing trajectories of massive test particles for ansatz~(\ref{Ca}) with $a=9$. \textbf{Left panel} ($J=2$): $E^2=0.89$, $\Delta\varphi=-2.44$ (red line), $E^2=0.93$, $\Delta\varphi=-2.76$ (green line), ${E^2=0.95}$, $\Delta\varphi=-2.89$ (blue line). The~dashed parts of lines represent the parts of trajectories located in the region $r<0$. \textbf{Right panel} ($J=3$): $E^2=0.925$, $\Delta\varphi=-0.82$ (red line), $E^2=0.965$, $\Delta\varphi=-0.64$ (blue line).}
  \label{F3}
\end{figure}

The possible shapes of orbits are represented in~Figures~\ref{F2}~and~\ref{F3}. Note that in this, and~in the next subsections, we represent all the orbit trajectories in the polar Schwarzschild-like coordinates, that~is, as~functions $C(\varphi)$ but not $r(\varphi)$. We see that there exist two distinct kinds of orbits: orbits of the first kind (second kind) are shown in the left panels (respectively, right panels) of these figures. Orbits of the first kind cross the wormhole throat, so that the parts of their trajectories that are located in the region $r<0$, and~shown by dashed lines in the figures, should be considered as hidden behind the throat. Therefore, for~such orbits, the~angle of precession, $\Delta\varphi$, of~an orbit should be determined observationally as the angular size of the observed trajectory of the orbit (that is, the~angle between an entry point in the region $r>0$ and the next exit point) minus $2\pi$; in other words, it should be set~$r_{p}=0$ for the lower limit of integration in~(\ref{precess}). It is obvious from the equality~(\ref{rJ}) and is shown in the left panel of~Figure~\ref{F1} that any wormhole of the first type has its own value $J_0$ of the specific angular moment such that orbits, independently of the specific energy $E$, are of the first kind if $J<J_0$; for $J>J_0$, an~orbit is of the first kind and of the second kind if, respectively, $V_{eff}(0,J)\leq{}E^2<1$ and $V_{eff}(r_*(J),J)<E^2<V_{eff}(0,J)$, where $r_*(J)$ is determined by Equation~(\ref{r*J}); orbits with $E\geq1$ are not bound, since $V_{eff}\rightarrow1$ as $r\rightarrow\infty$.

\unskip

In a wormhole of the second type, the~classification of bound orbits, as~can be seen directly from~Figure~\ref{F1} (right panel), depends on whether $0<J<J_*$ or $J>J_*$, where $J_*$ is determined by the expression~(\ref{J*}). Let the first condition be true. Then an orbit is of the first kind if $V_{eff}(0,J)\leq{}E^2<1$; an orbit is of the second kind if $V_{eff}(r_*(J),J)\leq{}E^2<V_{eff}(0,J)$. If~$J>J_*$ and $V_{eff}(r_*(J),J)\leq{}E^2<1$, then an orbit is always of the second kind. In~other cases bound orbits in a wormhole of the second type do not exist. Note that $J_0$, $J_*$, and~$r_{min}$ are determined only by a given wormhole spacetime (in our case, by~the function $C(r)$), while $r_*$ depends on $J$. However, this fact is not important in our analysis because there are no orbits with $E^2<V_{eff}(r_*(J),J)$.

For both the types of wormholes under consideration, the~angle of precession of orbits of the first kind decreases and attains negative values with increasing the specific energy $E$ when the value of~$J$ is fixed. Orbits of the second kind are completely located in the 'right' halves of wormholes. The~angle of precession of such an orbit, in~contrast to orbits of the first kind, increases with increasing the specific energy $E$ and at a fixed $J$. It also turns out that for orbits of the second kind, $\Delta\varphi$ is always positive in wormholes of the first type; but in wormholes of the second type, $\Delta\varphi<0$ for orbits that pass sufficiently close to the throat. The~various numerical experiments with wormhole solutions allow us to conclude that these properties hold generally and therefore can be used in astronomical observations in order to distinguish scalar field wormholes from scalar field naked singularities~\cite{Potashov2019} and scalar field black holes~\cite{Potashov2019a}.

It can be seen directly from the left panel in~Figure~\ref{F1} that wormholes of the first type can have a continuous accretion disk with the observable innermost stable circular orbit whose radius equals the radius of the throat $C_0=C(0)$. In~the first example ($a=1$) we have $C_0=4$, $0\leq{}J\lesssim3.82$, and~$4.97\lesssim{}E^2\lesssim9.61$. In~the innermost stable circular orbit, particles are expected to have a maximum possible specific angular momentum, because~any such particle has spiraled into this orbit from an outer orbit as a result of the loss of its energy and angular momentum: in the first example, $J_{ISCO}\approx3.82$ and $E_{ISCO}^2\approx9.61$. It can also be expected that the throat (in this context, the~visible inner edge of the accretion disk) appears to be bright enough, since the motion of particles in the neighbourhood of the throat should be slightly chaotic, but~enough to form a turbulent~flow.

In contrast, a~wormhole of the second type can be characterized by the condition $r_{min}=r_*(0)$, where $r_*(0)$ is the unique solution of the equation $V_{eff}'(r,0)= 0$ (or, equivalently, $A'(r)= 0$). Therefore,~ the~wormhole does not have the innermost stable circular orbit, but~instead has the resting-state stable circular orbit whose radius, $C(r_{min})$, always greater than $C_0$ (see last but one paragraph in Section~\ref{TimelikeNullG-1}). In~a sufficiently small neighborhood of the sphere of radius $r_{min}$, a~particle can be at rest at all times in the resting-state stable circular orbit or can slowly move near it in an orbit of small eccentricity. Therefore, the~wormhole will have a classical-like accretion disk if the baryonic matter surrounding the wormhole had been accreting onto it for a sufficiently short period of cosmic time before observations. In~this case, the~luminosity of the disk in the vicinity of its inner edge increases with increasing $r$. The~particles will spread out with time from the inner edge into a thin spherical shell of radius $r_{min}$. Thus, for~an 'old' wormhole with high accretion activity in the outer region, one can expect to see such a shell together with the adjoining inner edge of the accretion disk; both will be of small but nonzero brightness and will consist of 'grey' dust, gas, or~fluid.

\subsection{Trajectories of Null~Geodesics}

\begin{figure}[h]
  \centering
  \begin{minipage}{0.45\textwidth}
    \includegraphics[width=\textwidth]{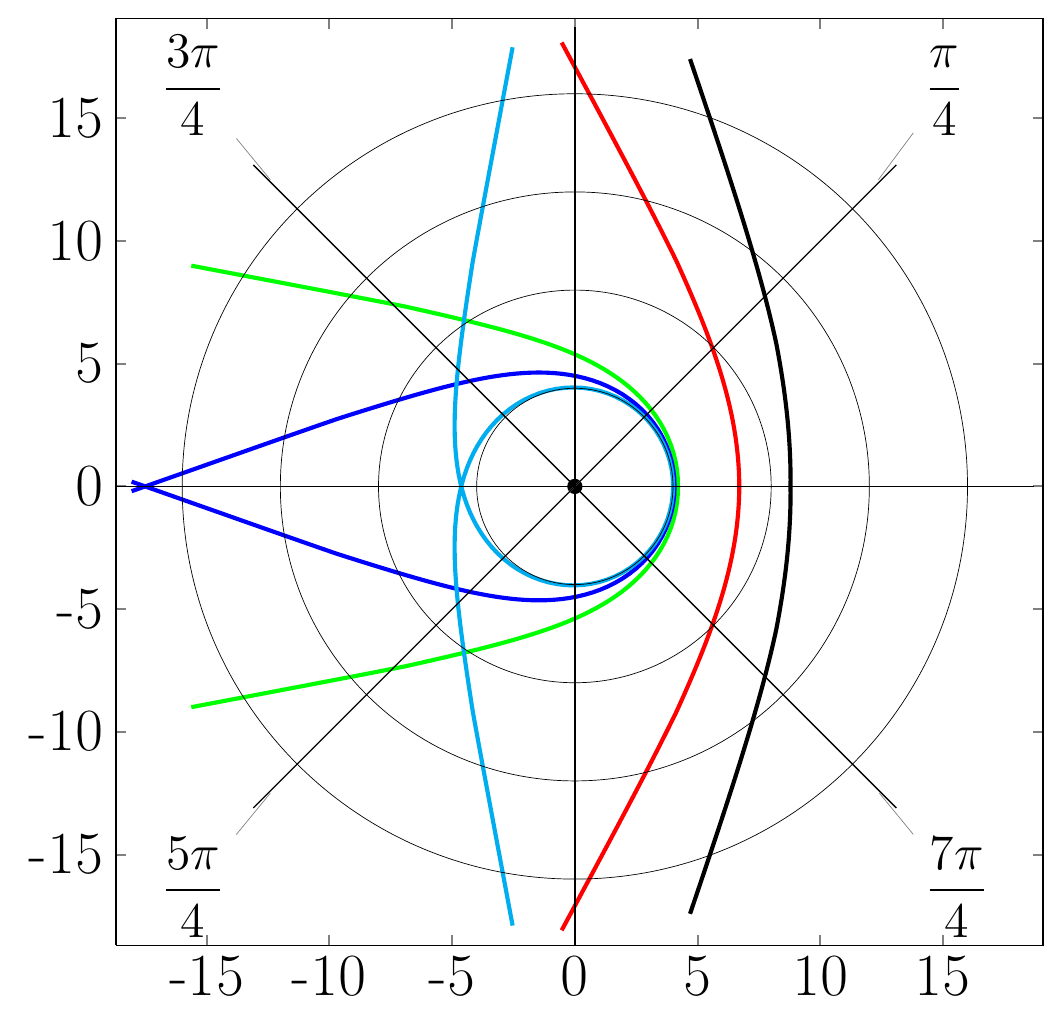}
  \end{minipage}\quad
\begin{minipage}{0.45\textwidth}
    \includegraphics[width=\textwidth]{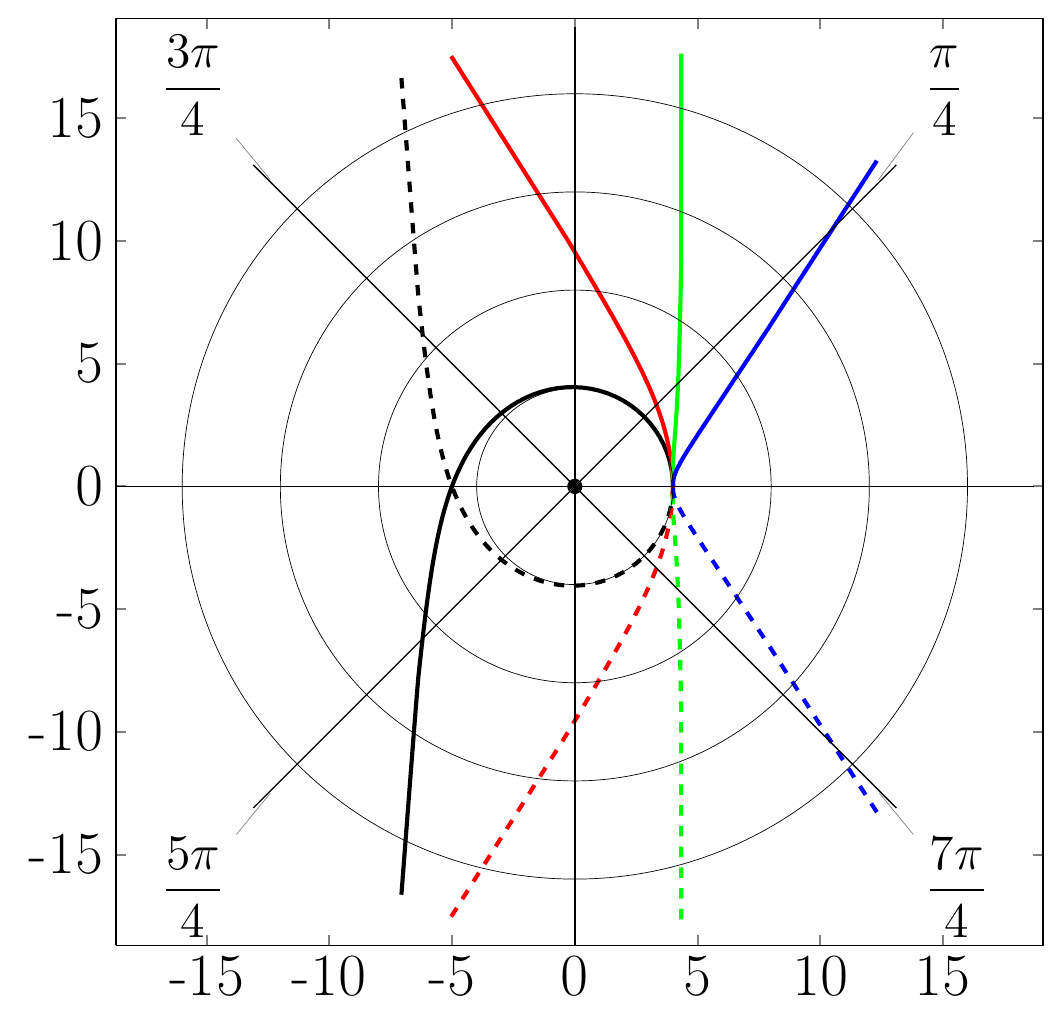}
  \end{minipage}
  \caption{Photon trajectories of massive test particles for ansatz~(\ref{Ca}) with $a=1$. In~the \textbf{left} plot: $b=10$~(black line), $b=8$~(red line), $b=5.8$~(green line), $b=5.7$~(blue line), $b=5.66$~(cyan line). In~the \textbf{right} plot: $b=5.65$~(black line), $b=5$~(red line), $b=4.2$~(green line), $b=3$~(blue line). The~dashed parts of lines represent the parts of trajectories located in the region $r<0$.}
  \label{F4}
\end{figure}

\begin{figure}[h]
  \centering
  \begin{minipage}{0.45\textwidth}
    \includegraphics[width=\textwidth]{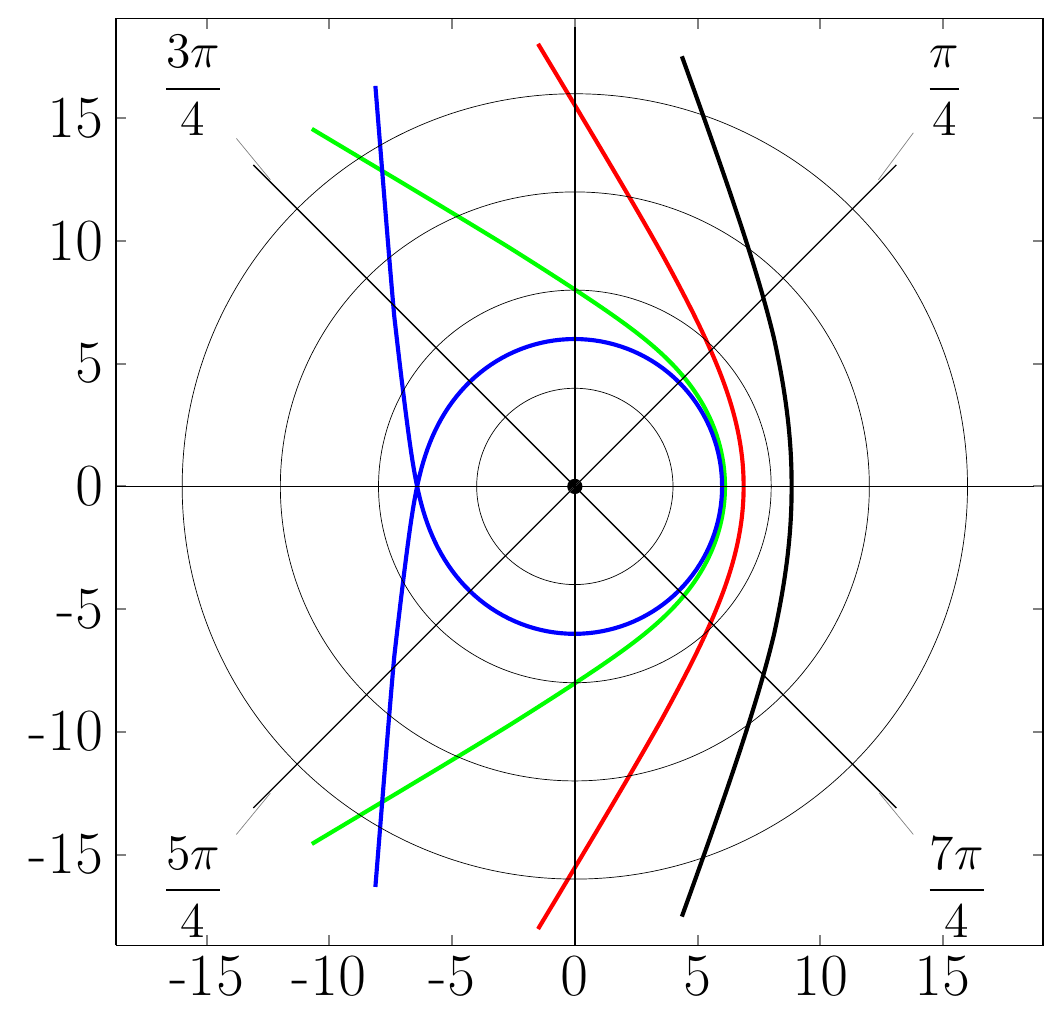}
  \end{minipage}\quad
\begin{minipage}{0.45\textwidth}
    \includegraphics[width=\textwidth]{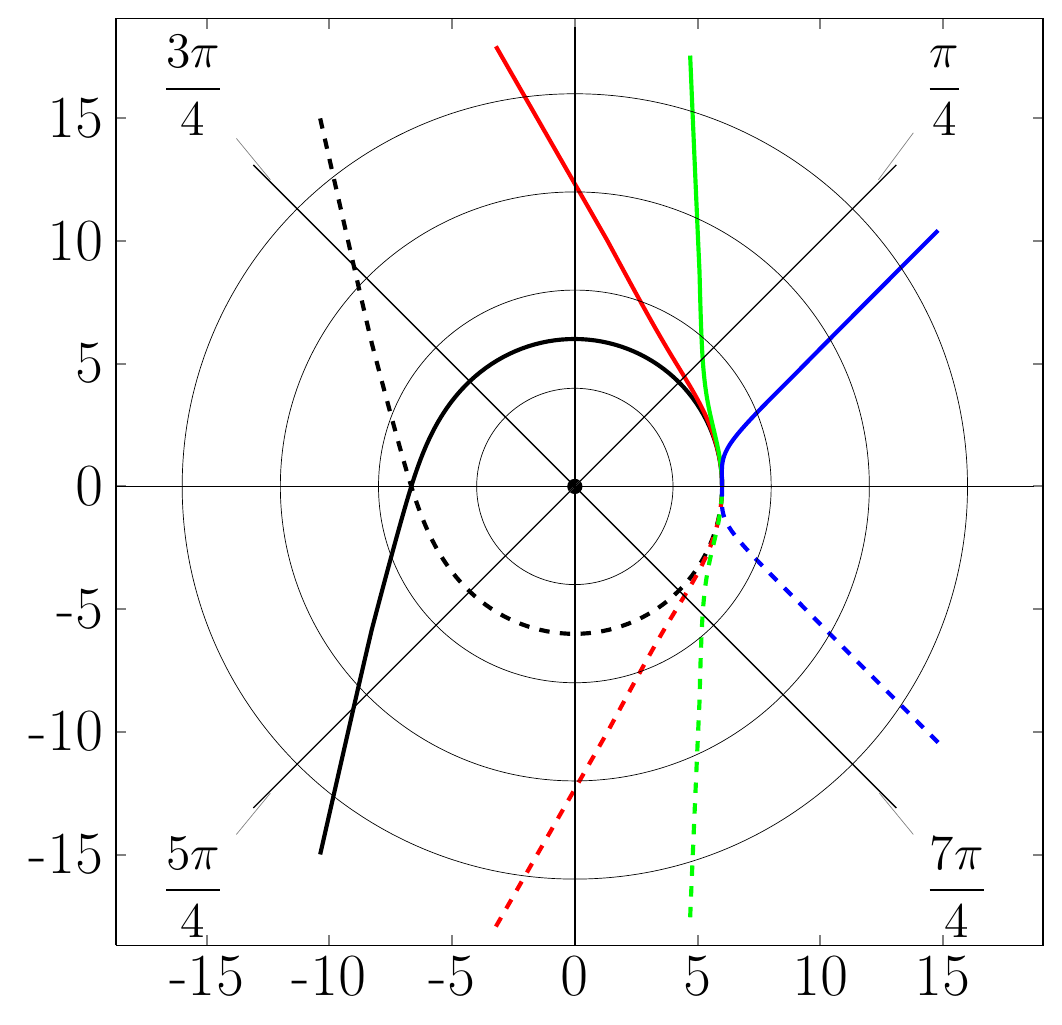}
  \end{minipage}
  \caption{Photon trajectories of massive test particles for ansatz~(\ref{Ca}) with $a=9$. In~the \textbf{left}~plot: \mbox{$b=10$~(black line)}, $b=8$~(red line), $b7$~(green line), $b=6.7$~(blue line). In~the \textbf{right} plot: \mbox{$b=6.65$~(black~line)}, $b=6$~(red line), $b=5.3$~(green line), $b=3$~(blue line). The~dashed parts of lines represent the parts of trajectories located in the region $r<0$.}
  \label{F5}
\end{figure}

Similarly to the case of massive test particles, there exist two kinds of null geodesics in a wormhole spacetime: a geodesic of the first kind passes through the wormhole throat, while a geodesic of the second kind remains in the region $r\geq0$. A~null geodesic is of the first kind if and only if the condition~(\ref{leave}) holds. Geodesics of the first kind (second kind) are shown in the left panel (respectively,~right panel) of Figures~\ref{F4}~and~\ref{F5}.

We will assume that the apparent luminosity of sources, placed at large distance from the wormhole, is very small. Then the image of a wormhole, that is, the~shape and brightness of the wormhole accretion disk, is determined by the behaviour of null geodesics of both these kinds, together with the inclination of the disk plane with respect to the line of sight of an observer. In~general, an~analytical computation of this image for scalar field wormholes seems to be impossible, therefore we will briefly discuss these issues qualitatively in the same standard manner as for black holes~\cite{Johannsen2016, Shaikh2018, Nandi2006, Mishra2018, Gourgoulhon2018, Dokuchaev2019}.

\unskip

If a wormhole of the first type with high accretion activity has the inner edge of the accretion disk to be placed at the throat, then an observer will see also the dual part of the disk placed in the region $r<0$. This means that the observer will see a continuous disk without a central dark spot; it~can be seen directly from Figure~\ref{F4} (right panel). As~it is shown above, a~wormhole of the second type with high accretion activity has a grey spherical shell of the observed radius $C(r_{min})$. Thus, in~fact, the~wormholes of interest are those that exhibit a low accretion activity and have a sufficiently large radius of the visible inner edge of the accretion disk. In~this case, the~vicinity of the disk near the visible inner edge should be bright (should be grey) in wormholes of the first type (respectively, of~the second type). This situation is illustrated in the left panels of Figures~\ref{F4}~and~\ref{F5} for both the first and second types of~wormholes.

One way to represent trajectories of photons coming from the left half of a wormhole (and,~consequently, the~image of bright matter located in the left half) consists in the following. A~static wormhole has the topology $\mathbb{R}\times\sum$, where $\mathbb{R}$ is 'time', and~the 'space' $\sum$ is the connected sum of two copies of $\mathbb{R}^3\backslash\mathbb{B}^3_{C_0}$; the boundary of the ball $\mathbb{B}^3_{C_0}$ is obviously the throat of the wormhole. It is convenient to immerse the wormhole into $\mathbb{R}^3$ with spherical coordinates $(\tilde{r},\theta,\varphi)$: one cuts out the central ball $\mathbb{B}^3_{C_0}$ from $\mathbb{R}^3$ and defines the immersion (it looks like a Riemann surface) explicitly by the map $(\pm{}r,\theta,\varphi)\mapsto (\tilde{r}-C_0,\theta,\varphi)$. If~a photon passes through the throat, then its trajectory in $\mathbb{R}^3\backslash\mathbb{B}^3_{C_0}$ consists of two parts which are located in the same plane and possess the reflection symmetry with respect to the perpendicular to the throat in the intersection~point.

Let us assume for simplicity that the line of sight of an observer is perpendicular to the disk plane. Then, it can be expected that a distant observer will see, first, the~outer accretion disk with the inner radius $C(r_{*}(J))$ (the first type,~$J>J_0$) or $C(r_{min})$ (the second type). Second, he will see a bright ring of the radius $C_0$ made up of photons in unstable circular orbits at the throat (the photon sphere). It~is also possible that the observer will also see the inner accretion disk (of the outer radius $C_m<C_0$), that~is, the~image of the dual part (placed in the region $r<0$) of the total accretion disk. In~any case, this~picture is radically different from that for black~holes.

\section{Discussion}\label{Disscussion}

In this article, we have examined two types of scalar field wormholes and found some distinctive features of timelike bound orbits located in the vicinities of the wormhole throats, and~of the trajectories of null geodesics. In~our approach, a~scalar field models dark matter surrounding the centres of normal galaxies. The~most important aspect of our results is that they can help to distinguish between wormholes and other possible types of strongly gravitating objects in galaxies using astronomical observations of orbital precessions, accretion disks, and~the 'images' of the central dark (or grey) regions. Here, for~definiteness, we compare wormholes with Schwarzschild black~holes.

We have seen that orbits that are sufficiently close to the centre of a wormhole exhibit considerable retrograde precession (the angle of pericentre precession is negative), while those that for the most part are in the region of Schwarzschild gravity have small negative or positive precession angles: this~can be seen from the comparison of the left and right panels in Figures~\ref{F2} and~\ref{F3}. Moreover,~the~effect of retrograde precession is more significant for wormholes of the second type, and, also, the~angle of pericentre precession increases with increasing the specific angular momentum $J$. However, the~modern astrometric accuracy is not enough to observe these effects confidently. For~example, the~orbits in right panels of Figures~\ref{F2} and~\ref{F3} have the pericentre distance of order of the size of the throat, namely $\approx$5 and $\approx$7 (in geometrical units with $M=1$), respectively. Note that Schwarzschild black holes always exhibit only the prograde relativistic precession of pericentres. On~the other hand, the~angle of pericentre precession for S-stars in the centre of our Galaxy can be measured today in the region $C\gtrsim2000$ ($\sim$100~AU for $M\sim4\cdot10^6\,M_\odot$). In~order to detect a meaningful deviation from the Schwarzschild precession, the~accuracy of the measurements should be at least an order of magnitude larger than what has been achieved~now.

Observations of the accretion disk of a wormhole seems to be more consistent with the present-day astronomical instrumentation, and~therefore to be more informative to compare with the observations of star orbits. In~this connection, it is necessary to mention the following three important points. First,~it~is well-known that the spot in the centre of image of a Schwarzschild black hole, known as the black hole shadow, has the observed radius $C_{sh}\approx5.2$ in geometrical units with $M=1$. On~the other hand, the~visible inner edge of the accretion disk in a wormhole spacetime can have an arbitrary value of the observed radius $C_d$ in the range $C_d\geq{}C_0$ or $C_d\geq{}C(r_{min})$ according to whether the wormhole is of the first or second type, respectively; thus, $C_d$ does not depend directly on the wormhole mass. Second, assuming that the background radiation of the sky is negligible in comparison with that of the accretion disk, this central spot should be dark. In~contrast, in~wormholes of the second type with high accretion activity, the~grey shell, surrounding the throat, is expected to be of small but nonzero brightness. Third, a~black hole is expected to have a bright ring of the observed radius $C_{ph}=3$ formed by photons moving in spiral orbits from the photon sphere to infinity. Wormholes of the second type with small accretion activity also can have such a photon ring, but~the brightness of the accretion disk in the vicinity of its inner edge should be very small.

\vspace{6pt}


\end{document}